\documentclass[twocolumn,showpacs,preprintnumbers,amsmath,amssymb, prb, reprint,longbibliography]{revtex4-1}
\usepackage[english]{babel}
\usepackage{amsmath}
\usepackage{stmaryrd}
\usepackage{txfonts}
\usepackage{amssymb}
\usepackage{mathrsfs}
\usepackage{graphicx}
\usepackage{dcolumn}
\usepackage{bm}
\usepackage{epsfig}
\usepackage{color}
\usepackage[colorlinks=true, linkcolor=blue, citecolor=blue, urlcolor=blue]{hyperref} 

\begin{document}

\title{Non-circular skyrmion and its anisotropic response in thin films of chiral magnets under tilted magnetic field}

\author{Shi-Zeng Lin}
\email{szl@lanl.gov}
\affiliation{Theoretical Division, Los Alamos National Laboratory, Los Alamos, New Mexico
87545, USA}

\author{Avadh Saxena} 

\affiliation{Theoretical Division, Los Alamos National Laboratory, Los Alamos, New Mexico
87545, USA}

\begin{abstract}
We study the equilibrium and dynamical properties of skyrmions in thin films of chiral magnets with oblique magnetic field. The shape of an individual skyrmion is non-circular and the skyrmion density decreases with the tilt angle from the normal of films. As a result, the interaction between two skyrmions depends on the relative angle between them in addition to their separation. The triangular lattice of skyrmions under a perpendicular magnetic field is distorted into a centered rectangular lattice for a tilted magnetic field. For a low skyrmion density, skyrmions form a chain like structure. The dynamical response of the non-circular skyrmions depends on the direction of external currents.
\end{abstract}
\date{\today}
\maketitle

A magnetic skyrmion is a topologically protected spin texture, which has been observed in magnets without inversion symmetry recently such as MnSi and FeGe.~\cite{Muhlbauer2009, Yu2010a,Yu2011,Seki2012,Adams2012} A skyrmion is characterized by a topological charge $Q=\frac{1}{4\pi}\int dr^2 \mathbf{S}\cdot( \partial_x \mathbf{S}\times \partial_y \mathbf{S})=\pm1$ with $\mathbf{S}(\mathbf{r})$ being a unit vector describing the direction of the spin. The typical size of skyrmion is about 5 nm to 100 nm and skyrmions form a triangular lattice. In bulk crystal, the skyrmion lattice is stabilized in a small region close to the critical temperature in the temperature - magnetic field phase diagram.~\cite{Muhlbauer2009} Skyrmions are found to be much more stable in thin films.~\cite{Yu2010a,Yu2011} Skyrmions respond to various external stimuli, such as magnetic field, electric current and temperature gradient. One extremely attractive feature of skyrmions is that they can be depinned by a low current density of the order of $10^6\ \mathrm{A/m^2}$, which is  5 to 6 orders of magnitude smaller than that for magnetic domain walls.~\cite{Jonietz2010,Yu2012,Schulz2012} Moreover, the conduction electrons in a metal interact with skyrmions and acquire a Berry phase, which produces an emergent electromagnetic field acting on these electrons. This gives rise to the topological Hall effect which has been observed experimentally.~\cite{Neubauer2009,NagaosaRMP2010,Zang11} Skyrmions in insulators can be driven by a temperature gradient~\cite{Kong2013,Lin2014PRL,Mochizuki2014} or electric field~\cite{White2012,PhysRevLett.113.107203}. For their unique physical properties, skyrmions are believed to be a prime candidate for the next generation spintronic devices.~\cite{Fert2013,NagaosaTokura2013}

It is crucial to tailor the skyrmion structure to optimize the desired functionalities. For instance, to achieve high density memory utilizing skyrmions, one needs to have skyrmions with the size in the nanometer range. The size of the skyrmion can be controlled by spin anisotropy or external magnetic fields, while the density of skyrmions can be tuned by external magnetic fields. Even the chirality of the skyrmion can be tuned by the sign of the Dzyaloshinskii-Moriya (DM) interaction through chemical substitution.~\cite{Shibata2013} The skyrmion in these cases has circular shape and the response is isotropic. It is of fundamental interest and of relevance for applications whether there exist non-circular skyrmions with an anisotropic interaction between them. We note that such non-circular skyrmions were observed experimentally in strained crystals, where the DM vector becomes anisotropic.~\cite{Shibata2015} Here we propose a simple way to stabilize non-circular skyrmions in thin films of chiral magnet by tilted magnetic fields.

\begin{figure*}[t]
\psfig{figure=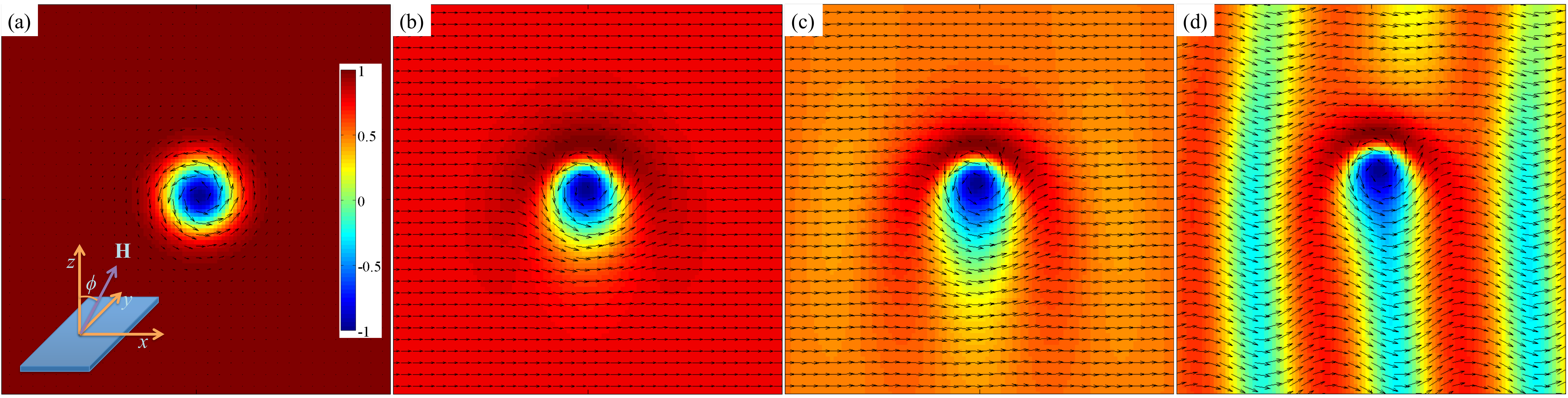,width=18cm}
\caption{(Color online) Skyrmion becomes non-circular when the magnetic field in the $x$-$z$ plane is tilted away from the normal of the film. The tilt angle $\phi$ is: (a)  $\phi=0^\circ$, (b) $\phi=40^\circ$, (c) $\phi=60^\circ$ (d) $\phi=66^\circ$. The color denotes the spin component along the $z$ direction, while the arrows represent the in-plane spin component. Inset in (a) is a schematic view of the setup. Here $H=0.8 D^2/J$ and the plotted area is $20\times 20 J^2 a^2/D^2$, with $a$ being the lattice constant of the spin system.
} \label{f1}
\end{figure*}

Skyrmions are more stable in thin films by suppressing the competing conical phase when a magnetic field is applied perpendicular to the thin film. When the magnetic field is tilted from the normal, the skyrmion phase is suppressed, which causes the decrease of the topological Hall resistivity as was measured recently in experiments.~\cite{PhysRevB.89.064416} When the field is parallel to the film, the skyrmion phase is suppressed completely and the conical phase is stabilized. Meanwhile the skyrmion shape is distorted for a tilted magnetic field because the region with spin parallel to the in-plane component of the field grows while the region with the opposite spin shrinks. Because of the distortion of skyrmion shape, the pairwise interaction between two skyrmions also becomes anisotropic, i.e. the interaction energy depends on the relative angle between the two skyrmions. The resulting skyrmion lattice is no longer a triangular lattice. For a low density of skyrmions, chains of skyrmions are stabilized because of the anisotropic interaction. At last, we will show that the dynamical response of skyrmions to an external current drive also becomes anisotropic.

We consider the following Hamiltonian for a classical spin $\mathrm{S}$ with $|S|=1$ in two dimensions ($x$-$y$ plane)~\cite{Yi09}
\begin{equation}\label{eq1}
\mathcal{H}=-J\sum_{\langle i, j\rangle}\mathbf{S}_i\cdot\mathbf{S}_j-D\sum_{i, \mu=x, y}\left( \mathbf{S}_i\times \mathbf{S}_{i+\mu}\cdot\mathbf{e}_\mu\right)-\mathbf{H}\cdot\sum_i \mathbf{S}_i,
\end{equation} 
where $J$ is the exchange interaction between the nearest neighbor spins, $D$ is the DM vector along the bond due to the breaking of inversion symmetry, $\mathbf{H}$ is the external magnetic field with a tilt angle $\phi$ from the $z$ axis, i.e. $H_x=H\sin\phi$ and $H_z=H\cos\phi$. Here $\mathbf{e}_\mu$ with $\mu=x,\ y$ is a unit vector in the $x$ and $y$ direction, respectively. The magnetic dipolar interaction was neglected because its strength is much weaker than the interactions in Eq. \eqref{eq1}. Equation \eqref{eq1} can reproduce satisfactorily the measured phase diagram in experiments. To obtain the skyrmion configuration at zero temperature $T=0$, we anneal the system using the Landau-Lifshitz-Gilbert equation with the Slonczewski 
spin-transfer torque term~\cite{Slonczewski1996}
\begin{equation}\label{eq2}
	{\partial _t}{\bf{S}}_i =- \gamma {\bf{S}}_i \times ({{\bf{H}}_{\rm{eff}}+\tilde{\mathbf{H}}}) + \alpha \mathbf{S}_i\times  {\partial _t}{\bf{S}}_i-\frac{\hbar\gamma}{2e} {J}_{\mathrm{ext}}\mathbf{S}_i\times(\mathbf{S}_j\times\mathbf{S}_i).
\end{equation}
The effective field is $\mathbf{H}_{\rm{eff}}\equiv-\delta \mathcal{H}/\delta {\bf{S}}_i$ and $\tilde{\mathbf{H}}$ is the Gaussian noisy field. Here $\alpha$ is the Gilbert damping coefficient and $\gamma$ is the  gyromagnetic ratio. To discuss the skyrmion dynamics, we have also introduced the spin current $J_{\mathrm{ext}}$ to describe the adiabatic spin-transfer torque between the conduction electrons and localized spins.

Let us first consider a single skyrmion in the tilted magnetic field. The spin configuration of a skyrmion as a function of tilt angle $\phi$ is presented in Fig. \ref{f1}. For a perpendicular magnetic field, the skyrmion is centrosymmetric. As the magnetic field deviates from the normal of the film, the skyrmion elongates along the direction perpendicular to the in-plane component of the magnetic field due to the Zeeman interaction term. For a tilt angle $\phi>66^\circ$, a clear coexistence phase of the skyrmion and the conical state can be seen. Such a skyrmion remains metastable even for $\phi=90^\circ$ when we tilt the field continuously towards the in-plane direction, because of the topological protection. 

 \begin{figure}[b]
\psfig{figure=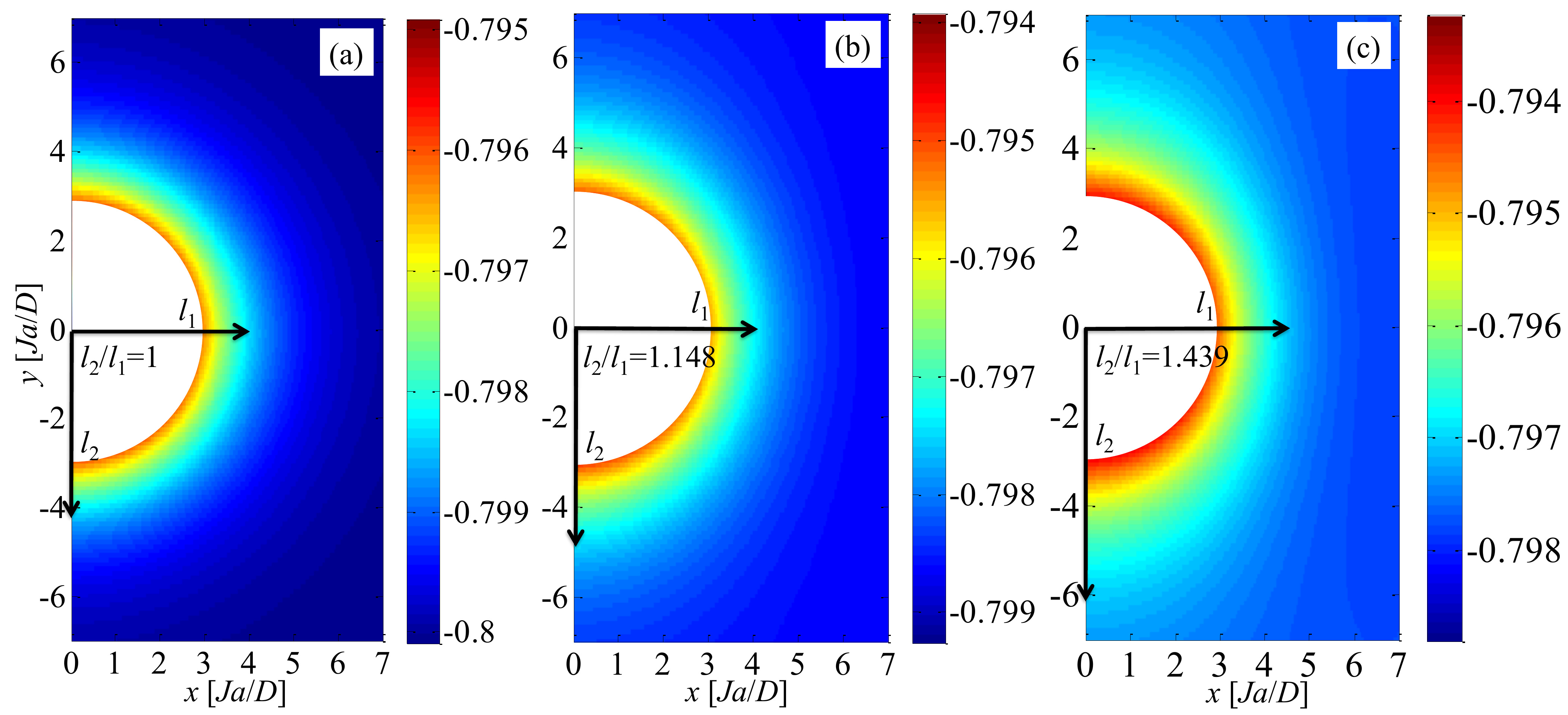,width=\columnwidth}
\caption{(Color online) Pair interaction $U(\mathrm{R}_1-\mathrm{R}_2)$ between two skyrmions at different tilt angle of the magnetic field: (a) $\phi=0^\circ$, (b) $\phi=45^\circ$ and $\phi=60^\circ$. Here $l_1$ and $l_2$ are two vectors along the principal axes connecting the origin and the energy contour at $U=-0.7974D^2/J$.  As indicated by the ratio $l_2/l_1>1$, the pair interaction becomes anisotropic when the magnetic field is tilted. Here $H=0.8 D^2/J$, $D=0.2J$, $T=0$ and $r_c=J a/D$. We restrict the pair interaction in the region $|\mathrm{R}_1-\mathrm{R}_2|\ge 3J a/D$ because the method to pin skyrmions by fixing spins in the skyrmion core becomes invalid when they are close to each other.
} \label{f2}
\end{figure}

For a non-circular skyrmion, the interaction between skyrmions becomes anisotropic. To calculate the pairwise interaction, let us first define the center of mass of a skyrmion as its topological center
\begin{equation}\label{eq3}
\mathbf{R}=\frac{1}{4\pi}\int d r^2 \mathbf{S}\cdot(\partial_x\mathbf{S}\times \partial_y\mathbf{S})\mathbf{r} .
\end{equation} 
Here we have used the continuum approximation, $\mathbf{S}_i\rightarrow \mathbf{S}(\mathbf{r})$, which is valid when $D/J\ll 1$. The interaction energy between two skyrmions, $U(\mathrm{R}_1-\mathrm{R}_2)$, depends on $\mathrm{R}_1-\mathrm{R}_2$, in contrast to the $|\mathrm{R}_1-\mathrm{R}_2|$ dependence for circular skyrmions. We then calculate $U(
\mathrm{R}_1-\mathrm{R}_2)$ by fixing spins of a skyrmion around its topological center $|\mathrm{r}-\mathrm{R}_1|<r_c$ in order to pin the skyrmion at a desired position. The results for $U(\mathrm{R}_1-\mathrm{R}_2)$ are shown in Fig. \ref{f2}. The interaction is mediated by exchange of magnons between two skyrmions and is repulsive. For circular skyrmions at a perpendicular magnetic field ($\phi=0$) the interaction is isotropic and is given by $U(\mathrm{R}_1-\mathrm{R}_2)=U(|\mathrm{R}_1-\mathrm{R}_2|)\sim K_0(|\mathrm{R}_1-\mathrm{R}_2|/\xi)$ for a large separation $|\mathrm{R}_1-\mathrm{R}_2|\gg J a/D$. Here the length scale $\xi$ is related to the magnon gap, $a$ is the lattice constant of the spin system and $K_0(x)$ is the modified Bessel function. For a tilted magnetic field, the interaction becomes anisotropic. The repulsion between skyrmions in the direction where they are elongated is stronger than that in the other directions.  This can be best seen by looking at $l_2/l_1$, where $l_1$ ($l_2$) is the vector along the principal axes connecting the origin and the energy contour at $U=-0.7974 D^2/J$. Here magnetic field has the component in the $ l_1$ direction. The ratio $l_2/l_1$ increases with the tilt angle $\phi$, meaning that the interaction becomes more anisotropic for a larger tilt angle.

\begin{figure*}[t]
\psfig{figure=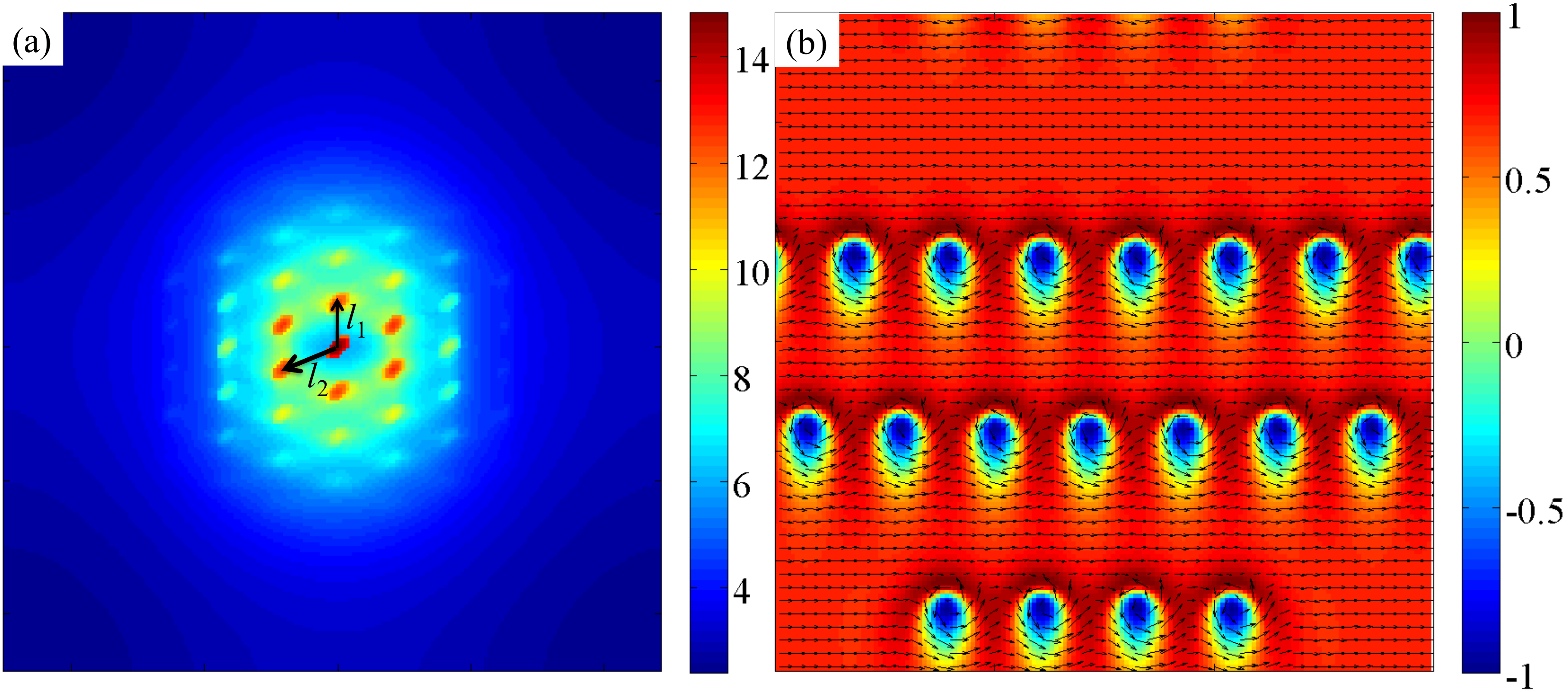, width=16.2cm}
\caption{(Color online) (a) Spin structure factor $\langle S_z(-\mathbf{q})S_z(\mathbf{q})\rangle$ plotted in the logarithmic scale at $H=0.336 J$, $D=0.7265J$, $T=0.05 J$ and $\phi=45^\circ$ obtained by Monte Carlo simulations of model Eq. \eqref{eq1} with $L=60a$. The skyrmions form a centered rectangular lattice with $l_1=0.0642\times 2\pi/a$ and $l_2=0.09016\times 2\pi/a$. (b)  A skyrmion chain state obtained by numerical annealing of model Eq. \eqref{eq2} to $T=0$ at $H=0.7 D^2/J$, $D=0.2J$, $\phi=48^\circ$ and $L=60 J a/D$. The color denotes the spin component along $z$ direction, while the arrows represent the in-plane spin component.} 
\label{f3}

\end{figure*}
\begin{figure}[b]
 	\psfig{figure=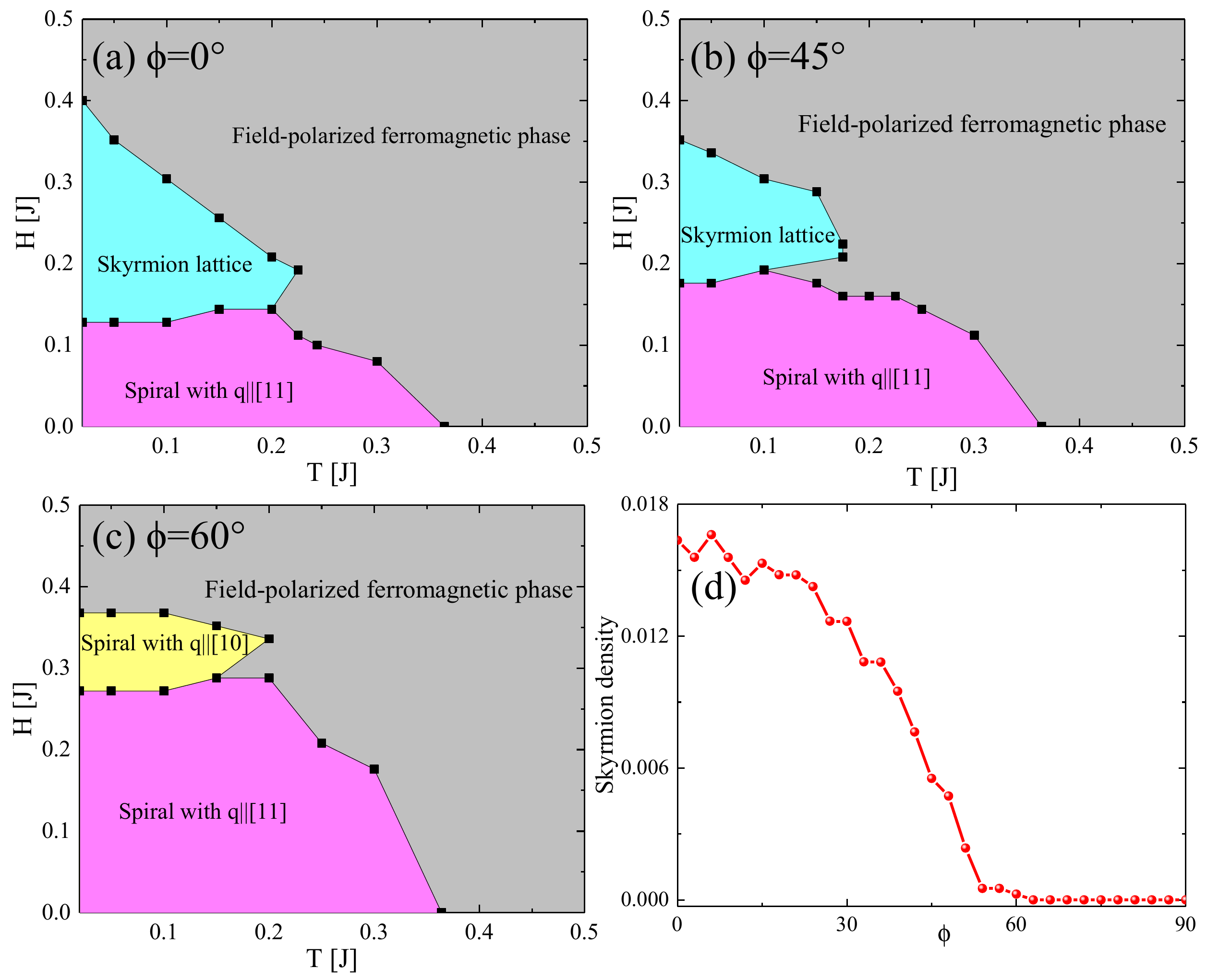, width=\columnwidth}
 	\caption{(Color online) (a-c) Temperature-magnetic field ($T$-$H$) phase diagram of the model Eq. \eqref{eq1} at different tilt angles of the applied magnetic field obtained by Monte Carlo simulations. Here $D=0.7265J$ and the system size is $L=60 a$. The phase boundary is obtained by checking the spin structure factor and spin configuration. (d) Skyrmion density as a function of tilt angle obtained by numerical annealing of the model from the paramagnetic state to $T=0$ at $H=0.7 D^2/J$ and $D=0.2J$.} \label{f4}
 \end{figure}
 
Such an anisotropic interaction between skyrmions has profound consequences for the equilibrium configuration of skyrmion lattice. The triangular lattice  is distorted into a centered rectangular lattice (space group p6mm to c2mm transition), i.e. the skyrmion lattice constant in the direction with a stronger repulsion becomes larger than that in the other directions. This can be seen from the spin structure factor $\langle S_z(-\mathbf{q}) S_z(\mathbf{q})\rangle$ displayed in Fig. \ref{f3} (a). Still we have six dominant Bragg peaks and higher order peaks for the skyrmion lattice. The ratio $l_2/l_1$ is larger than 1 because of the distortion due to the anisotropic interaction. In the coexisting phase of skyrmions and ferromagnetic state with a low skyrmion density, as shown in Fig. \ref{f3} (b), skyrmions arrange themselves into a chain along the direction with a weaker repulsion. We note that such skyrmion chains have also been observed in nanowires due to the geometry confinement. \cite{Tian2015}

Figures \ref{f4} (a-c) summarize the equilibrium phase diagram of model Eq. \eqref{eq1} at different tilt angles $\phi$. For a low tilt angle, we have a magnetic spiral with the ordering wavevector $\mathbf{q}$ along the $[11]$ direction at low magnetic fields. For intermediate magnetic fields, the skyrmion lattice was stabilized. The skyrmion lattice is distorted into a centered rectangular lattice for an oblique magnetic field. At high fields, we have a field-polarized ferromagnetic state. The skyrmion lattice phase shrinks with the tilt angle, and it disappears at $\phi\approx 60^\circ$. In this case the spiral with $\mathbf{q}$ along the $[11]$ direction transits into the conical phase with $\mathbf{q}$ parallel to the in-plane component of the magnetic field direction. (Here it is along the $x$ direction.) The skyrmion density as a function of the tilt angle $\phi$ at $T=0$ obtained by numerical annealing of Eq. \eqref{eq2} is presented in Fig. \ref{f4}(d).  The skyrmion density decreases and vanishes completely at $\phi\approx 60^\circ$. The threshold tilt angle where skyrmion density vanishes decreases with temperature. The decrease of the skyrmion density as evidenced from the topological Hall resistivity as a function of tilt angle was measured in $\mathrm{Mn_{0.96}Fe_{0.04}Si}$ thin films recently in Ref.~\onlinecite{PhysRevB.89.064416}, where the topological Hall resistivity vanishes at $\phi\approx 40^\circ$ at $T=20\ \mathrm{K}$. 

\begin{figure}[t]
\psfig{figure=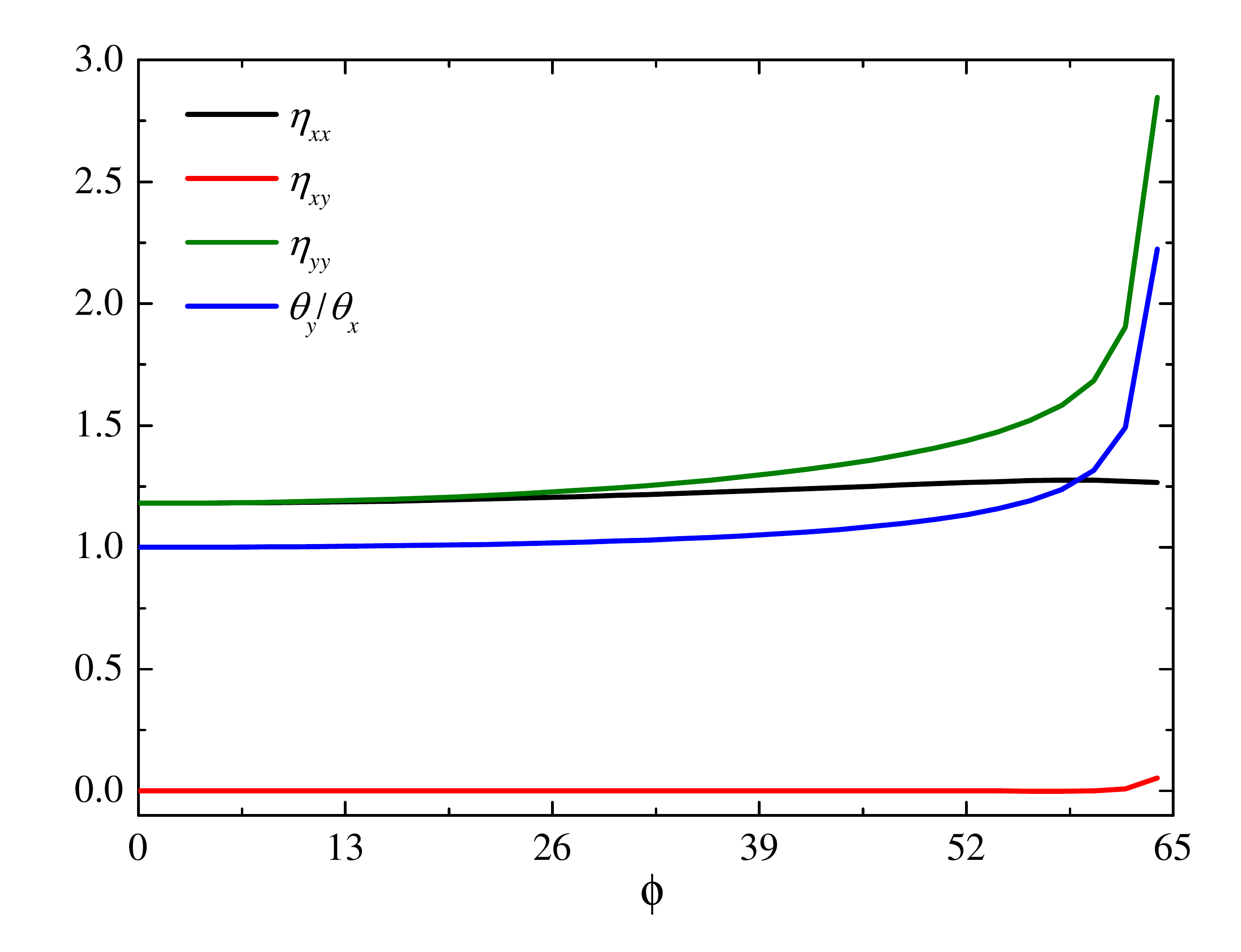, width=\columnwidth}
\caption{(Color online) Skyrmion form factors defined in Eq. \eqref{eq6} and the ratio of the Hall angle defined in Eqs. \eqref{eq7} and \eqref{eq8} for external currents in the $x$ and $y$ directions. Here $\alpha=0.1$, $H=0.8D^2/J$, $T=0$ and $D=0.2 J$. } 
\label{f5}
\end{figure}

Finally, let us discuss the equation of motion for the non-centrosymmetric skyrmion. It is more convenient to adopt the continuum approximation here. We follow Thiele's collective coordinate approach~\cite{Thiele72} by treating a skyrmion as a rigid object, i.e. $\mathbf{S}(\mathbf{r} ,t)=\mathbf{S}(\mathbf{r}-\mathbf{v} t)$. In this rigid skyrmion approximation, $\partial_t \mathbf{S}(\mathbf{r} ,t)\approx -(\mathbf{v}\cdot{\nabla}) \mathbf{S}$ and $\mathbf{S}\times\mathbf{H}_{\mathrm{eff}}=0$. After properly integrating out the internal degrees of freedom for skyrmions, we obtain the equation of motion for skyrmions as particles~\cite{Everschor12,szlin13skyrmion2,Iwasaki2013} from Eq. \eqref{eq2}
\begin{equation}\label{eq4}
\alpha  \eta_{ij} v_j-G_{ij} \left(J_{\mathrm{ext}, j}+v_j\right)=0.
\end{equation}
Here $i, j=x, y$ and summation over repeated indices is assumed. The tensor $G$ and the form factor tensor $\eta$ are given by
\begin{equation}\label{eq5}
G_{ij}=\frac{1}{4 \pi }\int dr^2 \mathbf{S}\cdot \left({\partial_i \mathbf{S}} \times{\partial_j \mathbf{S}}\right)=\left(
\begin{array}{cc}
 0 & -1 \\
 1 & 0 \\
\end{array}
\right),
\end{equation}
\begin{equation}\label{eq6}
\eta_{ij}=\frac{1}{4 \pi }\int dr^2 {\partial_i \mathbf{S}} \cdot{\partial_j \mathbf{S}}.
 \end{equation}
For a non-circular skyrmion, $\eta_{xx}\neq \eta_{yy}$ and $\eta_{xy}\neq 0$. We compute numerically $\eta_{ij}$ for skyrmions in Fig. \ref{f1} at different $\phi$, and the results are shown in Fig. \ref{f5}. While $\eta_{xy}\approx 0$ and $\eta_{xx}$ is almost independent on $\phi$, $\eta_{yy}$ increases rapidly with $\phi$ when the conical phase starts to appear. Note that $y$ is the direction along which skyrmions are elongated. For a current in the $x$ direction, the skyrmion acquires a velocity in the $y$ direction because of the damping. The Hall angle $\theta_x$ of the skyrmion motion is defined as
\begin{equation}\label{eq7}
\tan(\theta_x)=v_y/v_x=-\alpha\eta_{xx}/(1+\alpha\eta_{yy}).
\end{equation}
For a current in the $y$ direction, the Hall angle is given by
\begin{equation}\label{eq8}
\tan(\theta_y)=-v_x/v_y=-\alpha\eta_{yy}/(1-\alpha\eta_{xx}).
\end{equation}
As shown in Fig. \ref{f5}, the response of skyrmions to current becomes anisotropic, i.e. the Hall angle depends on the direction of the current, because of the non-centrosymmetric nature of the skyrmion in tilted magnetic fields.

To summarize, we have studied the equilibrium phase and dynamics of skyrmions with an oblique magnetic field. When the magnetic fields are tilted away from the normal of magnetic films, the skyrmion phase is less favorable and the skyrmion density decreases. Meanwhile, the shape of a skyrmion becomes non-centrosymmetric, which results in an anisotropic interaction between skyrmions. This anisotropic interaction stabilizes a centered rectangular lattice of skyrmions. In the low skyrmion density regime, a chain of skyrmions can be stabilized in the ferromagnetic background. Note that skyrmion chains have been observed in a confined geometry recently \cite{Tian}. The Hall angle of skyrmion motion depends on the direction of the current relative to the magnetic field direction. The predicted non-circular skyrmion and the resulting skyrmion configuration can be checked by imaging methods, such as Lorentz transmission electron microscopy or magnetic force microscopy.    

%\acknowledgments 

{\it Acknowledgments} Computer resources for numerical calculations were supported by the Institutional Computing Program at LANL. This work was carried out under the auspices of the NNSA of the US DOE at LANL under Contract No. DE-AC52-06NA25396, and was supported by the US Department of Energy, Office of Basic Energy Sciences, Division of Materials Sciences and Engineering.

%\bibliography{reference} 

\end{document}